\documentstyle[12pt]{article}
\newlength{\mytopmargin}
\newlength{\myleftmargin}
\begin{document}

\def\bsigma{\mbox{\boldmath$\sigma$}}
\def\half{\mbox{$\frac{1}{2}$}}

{\hfill {\bf cond-mat/9401001}}
\vspace{2mm}

\noindent
{\sl To appear in the Proceedings of the 16th.\ Taniguchi Symposium on
Condensed Matter Physics,
Kashikojima, Japan, October 26-29, 1993, edited by A. Okiji and
N. Kawakami (Springer, Berlin-Heidelberg-New York, 1994). }

\vspace{10mm}

\noindent
{\large
{\bf
Physics of the Ideal Semion Gas: Spinons and Quantum \\
Symmetries of the Integrable Haldane-Shastry Spin Chain.
}
\vspace{5.5mm}}

\normalsize
\noindent
F. D. M. Haldane
\vspace{5mm}

\noindent
Department of Physics, Princeton University, Princeton, New Jersey 08544, USA
\vspace{25.5mm}

\normalsize
\rm

\noindent
Various aspects of the Haldane-Shastry spin chain with $1/r^2$ exchange,
and its various generalizations, are reviewed, with emphasis on its
Yangian quantum group structure, and the interpretation of the model
as the generalization of an ideal gas (of ``spinons'') to the case of
{\it fractional statistics}.  Some recent results on its dynamical
correlation function are discussed, and  conjectured
extensions of these remarkably simple results to the $SU(n)$ model,
and to the
related Calogero-Sutherland model with integer coupling constant,
are presented.


\vspace{10mm}
\noindent
{\bf 1. Introduction}
\vspace{5mm}

\noindent
Most of our understanding of quantum many-body physics is based on
solvable models -- the ideal Bose and Fermi gases, and harmonic
oscillators.  These are the paradigms, solvable in full
detail, and treatments of more
complicated, interacting systems usually aim to find a description
of the system that is close to one of these paradigms.   There are very
few other fully-solvable models; the only general class of non-trivial
solvable
quantum models
are the integrable one-dimensional models that can be traced back to
the $S=1/2$ Heisenberg chain solved by Bethe \cite{bethe} in 1931.
Starting with Bethe, there has been success in calculating the
energy eigenvalues and thermodynamics of these models, and the algebraic
structures that make them solvable have been identified as the
quantum Yang-Baxter equation and ``quantum groups'', but progress in
explicit calculation of
their correlation functions  has so far been very limited.

Bethe's model and its generalizations generally involve {\it contact}
or {\it delta-function}
interactions.   A different class of
integrable models based on {\it inverse-square}
interactions was introduced around 1970 by Calogero \cite{calogero}
and
Sutherland  \cite{sutherland}.  However, this class of models
involved only gases of  spinless
impenetrable particles, and did not have the richness
and lattice generalizations of the Bethe family.  On the other hand,
certain (static) correlations functions were explicitly found
\cite{sutherland}.
A few years ago, spin-chain relatives of these
models were discovered
(the ``Haldane-Shastry'' model \cite{haldane88,shastry88}),
and it was subsequently found how to put
internal spin degrees of freedom
\cite{polychronakos,cherednik,hahaldane,wadati}
into the Calogero-Sutherland model (CSM).

In this paper, I will focus on the inverse-square-exchange spin-chain, and
its {\it trigonometric} $(1/\sin^2)$ variants \cite{haldane88,shastry88},
mentioning in passing also its {\it hyperbolic}
($1/\sinh^2)$ and {\it elliptic} $(1 /{\rm sn}^2)$
variants \cite{inozemtsev},
which provide a continuous interpolation to Bethe's model.
The trigonometric models are remarkable because they are explicitly
solvable in much
greater detail than Bethe's model, and share many of the characteristics
of the ideal gas, but with non-standard or {\it fractional} statistics,
which can occur in spatial dimensions below three.
In particular, it seems that the action of the physically-relevant
local operators (such as the spin operator on a given site of the
spin chain) on their ground states excites only a {\it finite} number
of elementary excitations.  This parallels the action of a one-particle
operator on an ideal gas ground state, and greatly simplifies the
calculation of correlation functions in terms of the ``form factors'',
the matrix elements of local operators between the ground state of
the system and eigenstates characterized by finite numbers of
elementary excitations.

In the {\it rational} limit (pure $1/r^2$ interactions), it has proved possible
to obtain simple explicit closed-form
expressions for the {\it thermodynamic potentials}
 \cite{haldane91a,sutherland93},
and very recently, for {\it ground-state dynamical correlation
functions}  \cite{haldane93}, which are again simple, but non-trivial.
I believe that these models will finally be solved in full detail,
approaching that with which the ideal gas can be solved, including,
for example, full correlation functions at all finite temperatures.
This should provide the first fully-developed extension to the standard
paradigms.

In this presentation, Section (2) will provide a self-contained
introduction to the Yangian ``quantum group''
\cite{drinfeld,haldane92,bernard93}
symmetries of these models; I will mainly restrict
the discussion to the $S$ = 1/2 spin-chain models with $SU(2)$
symmetry, and try to avoid the more esoteric mathematical
characterizations (Hopf algebras, coproducts, {\it etc.}),
disguising them in a more pedestrian physicists' terminology.
Section (3) describes the application of this to the eigenspectrum
of the trigonometric models, and Section (4) describes the recent
remarkably simple results \cite{haldane93}
for  some dynamical correlations of
the $S=1/2$ chain (and the $\lambda$ = 2 CSM), and presents new
conjectured (and certainly correct) generalizations to the
$SU(n)$ version of the chain, and to the CSM with arbitrary
integer coupling constant $\lambda$.

\vspace{10mm}
\noindent
{\bf 2. The Yangian ``Quantum Group'' and Integrable Heisenberg Chains}
\vspace{5mm}

\noindent
In this section, I will present a self-contained outline of the
Yangian ``quantum group'' algebra, and its
application to the Haldane-Shastry model (HSM) and its variants,
pointing out a number of open questions.  The HSM
contains a novel
realization of the Yangian  algebra, originally
characterized by Drinfeld \cite{drinfeld}
in the context of the algebraic Bethe Ansatz \cite{faddeev}, but
a mathematical structure in its own right.  I will not give a
comprehensive history of this elegant mathematical edifice;
a concise account  by Kirillov and Reshetikin \cite{kirillov}
references many of the original works.   An account of the representation
theory of the Yangian is given by Chari and Pressley \cite{chari}.

The original integrable model is the $S=1/2$ Heisenberg chain, solved
by Bethe in 1931 \cite{bethe}:
$$
H = \sum_i P_{i \, i+1} , \quad P_{ij} =
\half + 2 \vec{S}_i\cdot \vec{S}_{j} . \eqno (2.1)
$$
The integrability of
Bethe's model derives from  an underlying {\it ``quantum group''} algebra
called the $Y(gl_2)$ {\it Yangian} \cite{drinfeld}: let ${\bf T}(u)$ be a
$2 \times 2$ matrix with non-commuting operator-valued entries that
depend on a ``spectral parameter'' $u$, act in the Hilbert space
of the spin chain, and obey the algebra
$$
{\bf R}^{12}{\bf T}^1{\bf T}^2 =  {\bf T}^2{\bf T}^1{\bf R}^{12}  ,
\eqno (2.2)
$$
where ${\bf T}^1$ is the $4 \times 4$ operator-valued matrix
$({\bf T}(u_1) \otimes {\bf 1})$,  ${\bf T}^2$ is
$({\bf 1} \otimes {\bf T}(u_2))$, and
${\bf R}^{12}(u_1,u_2)$ is a $4 \times 4 $ c-number matrix defined
in the direct product space $V^1\otimes V^2$
of two $2\times 2$ c-number matrices.  Consistency of the algebra
requires that ${\bf R}$ satisfies
the {\it quantum Yang-Baxter equation} (QYBE)
${\bf R}^{12}{\bf R}^{13}{\bf R}^{23}$ =
${\bf R}^{23}{\bf R}^{13}{\bf R}^{12}$
in $V_1\otimes V_2 \otimes V_3$.
The Yangian $Y(gl_2)$ corresponds to the {\it rational} QYBE solution
$$
{\bf R}^{12} = (u_1-u_2){\bf 1} + h {\bf P}^{12} ;  \eqno (2.3)
$$
here ${\bf P}^{12}$ is the exchange matrix where (for c-number
matrices)
${\bf P}^{12}({\bf A} \otimes {\bf B} ) {\bf P}^{12}  $ =
$ ({\bf B} \otimes {\bf A})$, and
$h$ is the {\it quantum parameter}, which in this case sets
the scale  of the spectral parameter.

The Yangian algebra can be written as the commutation relation
$$
(u-v)[T^{\alpha\beta}(u),T^{\gamma\delta}(v)] =
h (T^{\gamma\beta}(v)T^{\alpha\delta}(u) -
T^{\gamma\beta}(u)T^{\alpha\delta}(v))  \eqno (2.4)
$$
It is commutative in the ``classical'' limit $h \rightarrow 0 $.
The  algebra implies that
$$
[t(u),t(v)] = 0 , \quad t(u) \equiv {\rm Tr}[{\bf T}(u) ] , \eqno (2.5)
$$
so the $t(u)$ are an infinite set of commuting operators.
Another consequence is that
$[{\bf T}(u), {\rm Det_Q}| T(v) | ]$ = 0 ,
where
${\rm Det_Q} |T (v) | $ is the {\it quantum determinant} \cite{korepin}
$$
{\rm Det_Q} | T(u) | =
T^{11}(u) T^{22}(u-h) - T^{12}(u)T^{21}(u- h) . \eqno (2.6)
$$
The quantum determinant
commutes with all elements of the algebra and is
analogous to
a Casimir operator.   In the ``classical
limit'' $h \rightarrow 0 $, the quantum determinant
reduces to the usual determinant of c-number-matrices.

It is consistent
to impose the asymptotic condition
${\bf T}(u)$ $ \rightarrow$ $ {\bf 1}$
for $ u$ $ \rightarrow$ $\infty $.
An asymptotic expansion can then be defined:
$$
{\bf T}(u) = \phi (u)
\left ( {\bf 1} + {h\over u} \left ( \vec{J}_0\cdot \vec{\bsigma} +
\sum_{n=1}^{\infty}
{ ( \vec{J}_n\cdot \vec{\bsigma} + J_{n}^0{\bf 1} )
\over u^{n} } \right )\right ) ; \quad
\phi (u) = \left ( 1 + \sum_{n=1}^{\infty}{ a_n \over  u^{n} }\right ) ,
\eqno (2.7)
$$
where $\vec{\bsigma}$ are Pauli matrices,
and  $\{a_n\}$ are the commuting generators of an infinite set of
Abelian subalgebras of $Y(gl_2)$:
$[a_m,a_n]$ =
$[\vec{J}_m,a_n]$ =
$[J_m^0,a_n]$ = 0; they alone determine the quantum determinant:
$$
\quad {\rm Det_Q} |T(u)| = \phi (u) \phi (u-h) . \eqno (2.8)
$$
The infinite-dimensional non-Abelian
$Y(sl_2)$ Yangian subalgebra, is completely generated by $\vec{J}_0$,
the generator of the $sl_2$ Lie algebra (of $SU(2)$ generators,
so $[J^a_0,J^b_n]$ = $i \epsilon^{abc} J^c_n$)
and the additional generator $\vec{J}_1$.
All other operators can be expressed in terms of these: for example,
$$
\quad J_1^0 = \half h \vec{J}_0\cdot \vec{J_0};
\quad J_2^0 = h \vec{J}_0 \cdot \vec{J_1} ;\quad  \ldots \quad ,\eqno (2.9a)
$$
$$
\vec{J}_2 = -i \vec{J}_1 \times \vec{J}_1 + \half h^2 (\vec{J}_0\cdot\vec{J}_0)
\vec{J}_0 ; \quad \ldots \quad . \eqno (2.9b)
$$
Note that, as a consequence of the Yang-Baxter relation, $[J_m^0,J_n^0]$ = 0;
also $[\vec{J}_0,J_n^0]$ = 0, but $[\vec{J}_1,J_n^0]$ $\ne$ 0.

The requirement that  ${\bf T}(u)$  obeys the $Y(gl_2)$ algebra
imposes consistency conditions on the $Y(sl_2)$ generators;
the first non-trivial
condition is the {\it Serre  relation}
$$
[J^a_1,J^b_2] = [J^a_2,J^b_1] . \eqno (2.10)
$$
In fact, this relation is sufficient to ensure that $\vec{J}_1$
generates $Y(sl_2)$, and can be used to compute the
value of $h^2$; either sign of $h$  = $\pm \surd h^2$
can be consistently chosen.
If the ``classical limit'' $h \rightarrow 0$
is taken, the $\vec{J}_n $ obey the infinite-dimensional
Lie algebra $\widehat{sl_2}_+$:
$$
[J^a_m,J^b_n] = i\epsilon^{abc} J^c_{m+n} , \quad m,n \ge 0 . \eqno (2.11)
$$
(A  familiar example of this algebra is the
subalgebra of any $SU(2)$
Kac-Moody algebra defined by its non-negative modes).

The {\it fundamental representation} of $Y(gl_2)$
(called the ``evaluation'' in the mathematical literature
\cite{drinfeld,chari}) is
$$
{\bf T }_1(u) = {\bf 1} +{ h{\bf P}_1 \over u  }; \quad
{\bf P}_1 = \half{\bf 1} + \vec{\bsigma }\cdot \vec{S}_1
\eqno (2.12)
$$
where $\vec{S}_1$ is a $S=1/2$ spin.
In this irreducible representation, $\vec{J}_1 $ = 0, and the
quantum determinant is  $(u+h)/u$.

If ${\bf T}(u)$ is a representation of $Y(gl_2)$, so is
$f(u){\bf T}(u - a)$, where $f(u)$ is a c-number function
($f(u) \rightarrow 1$ as $ u \rightarrow \infty$) and $a$ is a
spectral parameter shift.   The corresponding change in
$Y(sl_2)$ is $\vec{J}_1$ $\rightarrow$ $\vec{J}_1 + a \vec{J}_0$.

Like Lie algebras, ``quantum group''
algebras have a fundamental property
that allows larger representations
to be constructed from tensor-products of smaller representations
(the ``coproduct''); however, unlike Lie algebras, the
tensor-product operation is ``non-cocommutative'', which means that
the result of a tensor product depends on the {\it order} in
which it is carried out.   In the case of $Y(gl_2)$,
if ${\bf T}_1 (u) $ is a representation
acting on a Hilbert space ${\cal H}_1$ and ${\bf T}_2(u)$ is another
representation acting on another Hilbert space ${\cal H}_2$, then the matrix
product
$$ {\bf T}(u) = {\bf T}_1(u) {\bf T}_2(u) \eqno (2.13) $$
is also a representation, acting on a Hilbert space
${\cal H}_1\otimes {\cal H}_2$, and
$$
{\rm Det_Q} |T(u) | = ({\rm Det_Q}|T_1(u)|)( {\rm Det_Q}|T_2(u)|) .
\eqno (2.14)
$$
For the $Y(sl_2)$ subalgebra, the tensor product representation is
$$
\vec{J}_0 = \vec{J}_0(1) + \vec{J}_0(2) \eqno (2.15a)
$$
(the usual ``law of addition of angular momentum'') and
$$
\vec{J}_1 = \vec{J}_1(1) + \vec{J}_1(2) +
i h \vec{J}_0(1) \times \vec{J}_0(2). \eqno (2.15b)
$$
Note that reversing the ordering of $\vec{J}_0(1)$ and
$\vec{J}_0(2)$ defines an alternative tensor product with
$h \rightarrow -h$; either definition can be used, which
is why only $h^2$ is fixed by the Serre relation.  In $Y(gl_2)$
the two alternative tensor products are right and left
matrix multiplication.

A (Yangian) {\it highest-weight state} (YHWS), is a state that, for all $u$,
is annihilated by $T^{21}(u)$, and  is an eigenstate
of $T^{11}(u)$ and $T^{22}(u)$ with eigenvalues $\phi_1^+ (u)$ and
$\phi_2^+(u)$, where
$$
({\rm Det_Q}|T(u)|)|{\rm YHWS} \rangle =
\phi_1^+(u)\phi_2^+(u- h) | {\rm YHWS} \rangle. \eqno (2.16)
$$
Every finite-dimensional representation has at least one highest-weight
state.  It is {\it irreducible} if there is {\it only} one YHWS;
conversely,
an irreducible representation can be derived from each YHWS contained
in a reducible representation.
To each such YHWS  can be associated an
invariant subspace of all the states that can be generated from it
by successive action of elements of the algebra.
In contrast to what happens in the unitary representation theory
of  Lie groups and their associated Lie
algebras,  the representation of
$Y(gl_2)$ within
this invariant subspace may still be reducible, as the subspace
may  contain {\it other} YHWS (and their invariant subspaces)
which are eigenstates of the
quantum determinant with the same eigenvalue.

An irreducible representation containing (only) a given YHWS is obtained by
constructing its invariant subspace,  and then
projecting out the invariant
subspaces of all other YHWS contained within it.  This
is not necessary in the important special
case of {\it full reducibility}, associated with ``quantum symmetries'',
when the invariant subspaces associated with each independent YHWS
are always irreducible, and  orthogonal to each other.
Reducibility without {\it full} reducibility, where the YHWS $|2\rangle$
is contained within the invariant subspace generated by YHWS $|1 \rangle$,
means that there are non-vanishing matrix elements of the form
$\langle 2 | f({\bf T}(u)) | 1 \rangle $, but all matrix elements
of the form
$\langle 1 | f({\bf T}(u)) | 2  \rangle $
vanish.

The representation theory of $Y(sl_2)$ has been described in
detail in  \cite{chari}.
The basic theorem is that
all finite-dimensional irreducible representations are isomorphic to
an irreducible representation derived from the maximal-$J^z_0$ YHWS
of some tensor-product of fundamental representations.  This means that
the eigenvalues of $T^{11}(u)$ and $T^{22}(u)$ acting on a YHWS
which generates a finite-dimensional invariant subspace
have the form
$$
\phi_1^+(u) = f(u)P^+(u + h) ;\quad \phi_2^+(u) = f(u) P^+(u) ,
\eqno (2.17)
$$
where $P^+(u)$ is a finite-dimensional polynomial
called  the {\it Drinfeld polynomial}.
Let
$$
P^+(u)
= \prod_{n\ge 1}  \prod_{i=1}^{N_n} \left ( \prod_{\nu = 0}^{n-1} \left (
 u - u_i^{(n)} +\nu h  \right ) \right ),
\eqno (2.18)
$$
be the unique decomposition of the roots of $P^+(u)$ into
``strings'' of consecutive roots with spacing $h$,
obtained by successively factoring out strings, starting
with the longest ones present (this fixes how multiple roots are treated).
The dimension and $SU(2)$-representation content of the irreducible
representation of $Y(gl_2)$ derived from the YHWS
can immediately be identified from this form: it
is equivalent to a tensor product of $SU(2)$ representations where
$S=n/2$ occurs $N_n$ times, with a total dimension
$(2^{N_1})(3^{N_2})\ldots ((k+1)^{N_k})$, where the YHWS has
$J_0^z$ = $k/2$ = $\sum_n nN_n/2 $.
If only 1-strings are present in the Drinfeld
polynomial of a YHWS, the invariant subspace generated from it
must be irreducible.

The {\it Algebraic Bethe Ansatz} \cite{faddeev} is the solution to the problem
of finding the eigenvectors of  the commuting operators
$t(u)$ within an  irreducible finite-dimensional
representation of $Y(gl_2)$: the eigenstates  with eigenvalue
$t_{\nu}(u)$ which are
$SU(2)$ highest-weight states with $J^z_0$ = $(N/2) - M_{\nu}$, are given by
$$
|\nu \rangle =
\left (\prod_{i=1}^{M_{\nu}} T^{12} (v_i^{(\nu)})\right ) | {\rm YHWS} \rangle
\eqno (2.19)
$$
where the  $M^{(\nu)}$ ``rapidities'' $\{v_i^{(\nu)} \}$ are solutions
of the Bethe Ansatz equations (BAE)
$$
\phi^+_1(u) Q_{\nu}(u-h) +
\phi^+_2(u) Q_{\nu}(u+h)
= t_{\nu}(u) Q_{\nu}(u) \eqno (2.20)
$$
and $Q_{\nu}(u)$ is the polynomial
$$
Q_{\nu}(u) = \prod_{i=1}^{M_{\nu}} (u - v_i^{(\nu)} ) . \eqno (2.21)
$$
The eigenstates of $t(u)$ form an orthogonal basis only if
$[(\vec{J}_0\cdot\vec{J}_1),(\vec{J}_0\cdot \vec{J}_1)^{\dagger}]$ = 0.

The Bethe model on a periodic chain of $N$ sites with $S$ = 1/2 spins
corresponds to  a $Y(gl_2)$
representation that is a simple tensor product of fundamental
representations with no spectral parameter shifts:
$$
{\bf T}(u) =
{\bf T}_1(u)
{\bf T}_2(u)
\ldots
{\bf T}_N(u) . \eqno (2.22)
$$
It is irreducible, with $\phi^+_1(u)$ = $((u+h)/u)^N$ and
$\phi^+_2 (u)$  = 1.
The limit as $u \rightarrow 0 $ of $u^Nt(u)$ is
$h^N \exp i K $, where $\exp iK $ is the lattice translation
operator, and  we may write
$$
u^Nt(u) = h^Ne^{iK} \exp \left ( \sum_{n\ge 1} u^n H_{n+1} \right ) ,
\eqno (2.23)
$$
where $\{ H_{n} \}$ are a set of local Hamiltonians. Then
$[H_{m},H_{n}]$ = $[H_{m},\vec{J}_0]$ = $[H_m,\exp iK ]$ = 0,
and $H \propto H_{2}$.
The $Y(sl_2)$ generator may  be written
$$
\vec{J}_1 = i h  \sum_{i < j }  \vec{S}_i \times
\vec{S}_j . \eqno (2.24)
$$
In the conventional Bethe parameterization, $u$ is the rapidity
parameter if the choice $h$ = $i$ is made; in this case $\vec{J}_1$
is Hermitian.

For finite $N$, $[\exp iK, \vec{J}_1]$ $\ne$ 0, and
$[H_{m},\vec{J}_1]$ $ \ne$ 0.
However, in the {\it thermodynamic limit} $ N \rightarrow \infty $,
the model acquires a full non-Abelian {\it quantum symmetry}:
the Yangian generator $\vec{J}_1$
now {\it commutes} with the lattice translation
operator, and with all the $H_{n}$.
The infinite-dimensional
representation of the Yangian becomes {\it fully reducible}
into orthogonal subspaces in which the $H_{n}$ are diagonal.
Furthermore, the antiferromagnetic ground state becomes
the {\it unique} state that is singlet under the action of the Yangian.
In the  thermodynamic limit a spectral parameter ``boost''
operator $B$ \cite{tetelman}, where $[\vec{J}_1,B]$ = $h\vec{J_0}$,
is given by
$$
B = \sum_i (i + \half ) P_{i i+1}.   \eqno (2.25)
$$
This operator, which is incompatible with
periodic boundary conditions, also generates the family of
commuting Hamiltonians with quantum symmetry
from the lattice translation operator:
$[B,\exp iK]$  = $hH_2\exp iK$,
$[B,H_n]$ = $nhH_{n+1}$.

The most general representation of the $Y(sl_2)$ Yangian based on
a simple tensor product of fundamental representations (including
spectral parameter shifts) is
$$
\vec{J}_1 = \sum_i \gamma_i \vec{S}_i  + \half
\sum_{i \ne j} w_{ij} \ \vec{S}_i \times \vec{S}_j  \eqno (2.26)
$$
where $w_{ij}$ = ${ih\,\rm sgn}(i-j)$, and the $\{\gamma_i\}$ are arbitrary.
It is an instructive exercise to substitute a form of this type
into the Serre relation, {\it without} making any assumptions about
$w_{ij}$, except that it can generally be taken to be odd.
The result \cite{haldane92}
is that this is a representation of the Yangian generator if
$$
w_{ij}w_{jk} + w_{jk}w_{ki} + w_{ki}w_{ij} = h^2 , \quad i \ne j \ne k.
\eqno (2.27)
$$
The general solution of this is
$$
w_{ij} = ih \left ( { z_i + z_j \over z_i - z_j } \right )
\eqno (2.28)
$$
where $\{z_i\}$ are a set of $N$ arbitrary, but distinct, complex parameters.
The simple tensor-product representation is recovered in a limit
where $|z_i/z_j| \rightarrow \infty $ for $ i > j $.

Note that $Y(sl_2)$
representations with an invariant subspace where groups of $2S$ consecutive
$S$ = 1/2 spins remain in the symmetric spin-$S$ state can be obtained from
the simple tensor product by suitably choosing the associated
spectral parameter shifts $\gamma_i$
to be a $2S$-string, and this is the basis for the extension of
Bethe's model to chains of spins with $S > $ 1/2 \cite{takhtajan}.
However, this requires the stronger property $(w_{ij})^2  + h^2 $ = 0,
which eliminates  more general representations based on (2.26) and (2.28) for
such spins.

The  $Y(sl_2)$ representation based on (2.28) is evidently derived from
a more general class of representations of $Y(gl_2)$ than
those obtained by a simple tensor product.
Instead, following
Ref.\cite{bernard93}, they can
be obtained as follows.
Let $\{\gamma_i\}$ and $\{z_i\}$ be variables that commute
with each other, and with the spin variables, and
let $K_{ij}$ be a permutation operator that permutes their labels:
$K_{ij} \gamma_i$ = $\gamma_jK_{ij}$, {\it etc.}
These permutation operators do {\it not} affect the spins:
$[K_{ij},\vec{S}_k]$ = 0.
Now consider the quantities
$$
\hat{\gamma}_i = \gamma_i + h {\sum_{j (\ne  i)}}^{\prime}
\left ( {z_i \theta (j-i) + z_j \theta (i-j)
\over z_i - z_j} \right ) K_{ij} .
\eqno (2.29)
$$
These variables commute, and obey
a ``degenerate affine Hecke algebra'' \cite{bernard93}:
$$
K_{ii+1}\hat{\gamma}_i - \hat{\gamma}_{i+1}K_{ii+1} = h ;
\quad [\hat{\gamma}_i,\hat{\gamma}_j] = 0 .
\eqno (2.30)
$$
(The role of  Hecke algebras in this context
has been stressed in Ref.\cite{cherednik}.)
This implies that
$$
 [K_{ij}, \Delta (u) ] = 0, \quad \Delta (u)
\equiv  \prod_i (u-\hat{\gamma}_i),
\eqno (2.31)
$$
and hence that $\Delta (u)$ is a c-number function
of $u$ and the $\{\gamma_i,z_i\}$.
Since the $\hat{\gamma}_i$ commute with each
other and with the spins, a
representation of $Y(gl_2)$ is clearly given by
$$
{\bf T}(u) =  {\bf T}_1(u-\hat{\gamma}_1)\ldots
{\bf T}_N(u-\hat{\gamma_N}) ,
\eqno (2.32)
$$
where $\Delta (u){\bf T}(u)$ is a polynomial
of degree $N$ in $u$.

Now let $\Pi^+$ be a projection operator into the
subspace that is fully symmetric under simultaneous
permutation of spins and the parameters $\{\gamma_i,z_i\}$,
so $ K_{ij} \Pi^+$ = $ P_{ij}\Pi^+$.
The Hecke algebra guarantees  the property
$$
 \Pi^+ {\bf T}(u) \Pi^+
=  {\bf T}(u) \Pi^+.
\eqno (2.33)
$$
This ensures that the totally-symmetric subspace is an
invariant subspace of ${\bf T}(u)$, and hence that
the projection of (2.32) into this subspace
is {\it also} a representation of $Y(sl_2)$.
The representation
can now be evaluated \cite{bernard93}
within the totally-symmetric subspace as
$$
{\bf T}(u) = {\bf 1} + h \sum_{ij} {\bf P}_i ((u - L)^{-1})_{ij} ,
\eqno (2.34a)
$$
where  $L_{ij}$ are elements of a $N\times N$ ``quantum Lax matrix''
$$
L_{ij} = \gamma_i\delta_{ij} +
h (1-\delta_{ij})(1 - (z_i/z_j))^{-1}P_{ij} . \eqno (2.34b)
$$
The $2^N$ states of this representation are
all eigenstates
of
the quantum determinant with the same eigenvalue
$\Delta (u + h) /\Delta (u) $
where  $\Delta (u)$ is the polynomial of degree $N$  given by
$\det |u - L^0 |$ and
$L^0_{ij}$ is the eigenvalue of $L_{ij}$ acting on the fully-spin-symmetric
YHWS with $J^z_0$ = $N/2$
and $P_{ij}$ = 1.  This YHWS has the Drinfeld polynomial
$P^+(u)$ = $\Delta (u)$.

For generic values of the $\{\gamma_i\}$ and $\{z_i\}$, the representation
is irreducible.
We can now ask whether for some special choices of these parameters,
the representation becomes fully-reducible as a direct sum of smaller
representations, which would imply that there is a non-trivial
Hermitian
operator $H$ that
commutes with both $\vec{J}_0$ and $\vec{J}_1$.
We can  postulate such an operator to be of the
form $H $= $\sum_{i<j} h_{ij} P_{ij}$,
and look for solutions of the condition
$[H,\vec{J}_1] $ = 0.
This is satisfied provided
$$\gamma_i - \gamma_j = 0 ;\quad
h_{ij} \propto  (w_{ij}^2 + h^2);\quad
{\sum_{j (\ne i)}}^{\prime} w_{ij} h_{ij}  = 0 . \eqno (2.35)
$$
We also require that $H$ is Hermitian ($h_{ij}$ real).
We find \cite{haldane92} {\it two}
families of translationally-invariant
Hamiltonians with quantum symmetry, where
$[\exp iK, \vec{J}_1]$ = $[\exp {iK}, H ]$ = 0;
the Hamiltonian is even parity under spatial
reflection, while $\vec{J}_1$ is odd-parity, with
$(\vec{J}_1)^{\dagger} \propto \vec{J}_1$, which ensures
that any reducibility is {\it full} reducibility.
The solutions are
$\gamma_i$ = 0, and $h_{ij}$ $\propto$ $ 1/d(i-j)^2$, where
in the {\it hyperbolic models}, defined on an {\it infinite} chain,
$$
w_{ij} = i h \, \coth (\kappa (i-j)) ; \quad d(j) =
\kappa^{-1} \sinh (\kappa j ) \eqno (2.36a)
$$
with real $z_{i+1}/z_i$ = $\exp 2\kappa$, which can be either
positive ($\kappa $ real) or negative ($\kappa - i \pi/2  $ real).
In the limit $ \kappa \rightarrow \infty $, the hyperbolic model becomes
the $N = \infty $ Bethe model; in the limit $\kappa \rightarrow 0 $,
and $ih$ = $ 2\kappa$, so $ h \rightarrow 0 $, we get a ``classical
limit '' where $Y(sl_2)$ becomes the
infinite-dimensional  Lie algebra $\widehat{sl_2}_+$, and
we obtain the {\it rational model}
$$
w_{ij} = (i-j)^{-1} ;\quad  d(j) = j . \eqno (2.36b)
$$

The other family of solutions are the {\it trigonometric models},
obtained from the hyperbolic models by the replacement
$\kappa \rightarrow i\pi /N $, with
$$
  w_{ij} =  h  \cot (\pi (i-j)/N) ; \quad d(j) =
(N/\pi) \sin (\pi j/N ) , \eqno (2.37c)
$$
defined on a {\it finite} periodic chain on $N$ sites, with $d(j)$ being
the chord distance between lattice sites equally-spaced on
a circle. For the choice of real $h$,
the $N$ roots of $\Delta(u)$ form an
$N$-string along the real axis, centered on the origin.
The trigonometric model then has $(L_{ij})^{\dagger}$
= $L_{ji}$, and hence $(T^{\alpha\beta}(u))^{\dagger}$
= $T^{\beta\alpha}(u^*)$.

In all these models, $H$ $\propto$ $H_2$ is the first member of a
family of commuting constants of the motion
$\{  H_2 , H_3, H_4,\ldots \}$
that commute with $\vec{J}_0$, $\vec{J}_1$, and $\exp iK$.
The eigenstates of these operators form irreducible representations
of $Y(sl_2)$.
For example \cite{inozemtsev},
$$
H_3 \propto \sum_{i \ne j \ne k} {z_i z_j z_k \over
z_{ij} z_{jk} z_{kl} } \vec{S}_i \cdot \vec{S}_j \times \vec{S}_k ,
\quad z_{ij} \equiv z_i-z_j , \eqno (2.37)
$$
but the systematic construction beyond $H_4$ \cite{haldane92}
has not yet been
elucidated.
So far, no generalization of the spectral-parameter ``boost''
operator from the Bethe limit
to the hyperbolic and trigonometric
models has been found.
The origin of the Hamiltonian constants of the motion
in the models with quantum symmetry
is conceptually
rather different from that in the finite-$N$ Bethe model. They
must commute with all elements of ${\bf T}(u)$, but the only elements
of $Y(gl_2)$ with this property derive from the quantum determinant,
which is a trivial c-number in these representations. Thus
the Hamiltonians with
quantum symmetry {\it cannot be expressed  as functions of ${\bf T}(u)$},
and are independent objects.

The Yangian symmetry of
the {\it hyperbolic} models
is essentially similar to that of the Bethe model: it occurs in the
thermodynamic limit, and $\vec{J_1}$ is made Hermitian by the usual
BAE  choice
$h = i$. On the other hand,
the realization of Yangian symmetry in the {\it trigonometric}
models is essentially new, as it occurs in a {\it finite periodic }
system, which has discrete, highly-degenerate energy levels that
can be classified as finite-dimensional irreducible representations
of $Y(sl_2)$.  In this case $\vec{J}_1$ is made Hermitian by
choosing
$h$  {\it real}.  The two variants are only connected  via
the ``classical'' {\it rational} model with $h$ = 0.

The consequences of Yangian quantum symmetry
are very different in the two cases; there is an
analogy to the difference between the non-compact
Lorentz group in (1+1)-dimensions and $SU(2)$, which have
symmetry algebras
corresponding to different real forms of the same complex
$sl_2$ Lie algebra.
In the hyperbolic case, the representation theory involves the
Bethe ``rapidity strings'' with spacing $h$ = $i$ in the
imaginary rapidity direction, which are related to bound states.
In the trigonometric case, the string
spacing is in the real direction, and is related to the quantization
of the momentum of free particles with periodic boundary
conditions in units of $2 \pi /N $, and to  a
generalization of the Fock space structure of such a system.

The trigonometric models, and their rational limit, were independently
discovered by Haldane \cite{haldane88} and Shastry \cite{shastry88}.
The extension
to the hyperbolic model, was found by Inozemtsev \cite{inozemtsev},
who also proposed the integrability of an {\it elliptic} variant with
$h_{ij}$ $\propto$ $\wp (i-j)$, where $\wp (u)$ is the doubly-periodic
Weierstrass
function with periods $N$ and  $i\pi / \kappa $. This is a version of the
hyperbolic model, made periodic on a finite chain of $N$ sites, and
(like the finite-$N$ Bethe model ($\kappa \rightarrow \infty $)) does
{\it not} have quantum symmetry.  A proof of integrability
is missing, but two odd-parity operators that commute with the Hamiltonian
are given in
 \cite{inozemtsev}, making integrability very plausible.
One linear combination of these  corresponds to $H_3$
in the trigonometric and  hyperbolic limits, the other,
which corresponds to $\vec{J}_0\cdot \vec{J}_1$,
is (where $\zeta (u)$ is the elliptic zeta function)
$$
\sum_{i \ne j \ne k } \left ( \zeta (i-j) + \zeta (j-k) + \zeta (k-i) \right )
\vec{S}_i \times \vec{S}_j \cdot \vec{S}_k . \eqno (2.38)
$$
I have not been able to find a $Y(sl_2)$ generator
$\vec{J}_1$ where $\vec{J}_1\cdot \vec{J}_0$ gives (2.38).
I therefore conjecture that the integrability of
this model may involve a yet more general ``quantum group'' algebra
with ${\it two}$ ``quantum parameters'' $h$ = $2i\kappa$ and
$h'$ = $2\pi /N$ (which would be scale parameters for two distinct
spectral parameters), which only degenerates to the Yangian when either of
them vanishes or becomes infinite (the Bethe limit).
This would correspond to a ``double quantization''
of $\widehat{sl_2}_+$, or a further ``quantization'' of $Y(sl_2)$.
Indeed, Inozemtsev  \cite{inozemtsev}
suggests that two spectral parameters
play a role in this model.
A better understanding of the elliptic model
is clearly needed.

To conclude the formal discussion of algebraic aspects of these models,
I note that, as in the case of Bethe's model \cite{sutherland75},
there is a straightforward
extension from the $Y(gl_2)$ quantum group
with $SU(2)$ symmetry, to the $Y(gl_n)$ quantum group, where
${\bf T}(u)$ is an $n \times n$ matrix,
with $SU(n)$ symmetry. A further extension
is to the graded or {\it supersymmetric}
generalization, $Y(gl_{m|n})$, with $SU(m|n)$ supersymmetry,
where a site can be in one of
$m$ states with even fermion number or $n$ states with odd fermion
number.   Let $c^{\dagger}_{i\alpha}$ create on site $i$
a particle of species
$\alpha$, which may be fermionic or bosonic, and impose the constraint
$$
\sum_{\alpha} c^{\dagger}_{i\alpha}c_{i\alpha} = 1, \quad \mbox{ all } i.
\eqno (2.39)
$$
The exchange operator becomes
$$
P_{ij} = \sum_{\alpha\beta} c^{\dagger}_{i\alpha}c^{\dagger}_{j\beta}
c_{i\beta}c_{j\alpha},
\eqno (2.40)
$$
and the generators of the $sl_{m|n}$ graded Lie algebra and its
Yangian extension can be given as the
traceless operator-valued matrices
$$
J^{\alpha\beta}_0 = \sum_{i}  \left (
c^{\dagger}_{i\alpha}c_{i\beta} - (m+n)^{-1}\delta^{\alpha\beta} \right ),
\eqno (2.41a)
$$
$$
J^{\alpha\beta}_1 = \half \sum_{i\ne j, \gamma}
w_{ij} c^{\dagger}_{i\alpha}c^{\dagger}_{j\gamma}c_{i\gamma}c_{j\beta}.
\eqno (2.41b)
$$
In particular, the
graded trigonometric exchange model with $(m|n)$ = $(1|2)$ corresponds
to  the ``supersymmetric
$t$-$J$ model'' variant found by Kuramoto \cite{kuramoto}.

\vspace{10mm}
\noindent
{\bf 3.  Spectrum of the Trigonometric Haldane-Shastry Model,
and its Interpretation as a Generalized Ideal Gas}
\vspace{5mm}

\noindent

In this section I will focus on the remarkable properties of the
trigonometric model, with its realization of Yangian quantum symmetry
in a form compatible with periodic boundary conditions.
The Hamiltonian, normalized so the spin-wave velocity is $v_s$ in
units where the lattice spacing $a$ = $\hbar$ = 1, is
$$
H = {\pi v_s \over N^2}
 \sum_{i<j} { P_{ij} \over \sin^2 (\pi (i-j)/N) } . \eqno (3.1)
$$
Let us first consider the case where the symmetry group is
$SU(1|1)$, when the model is equivalent to a lattice gas
of free {\it spinless fermions}, with creation operators
$a^{\dagger}_i$ = $c^{\dagger}_{i2}c_{i1}$, so
$$
P_{ij} = 1 - (a^{\dagger}_i-a^{\dagger}_j)(a_i - a_j)
= -1 + (a_i-a_j)(a^{\dagger}_i-a^{\dagger}_j) . \eqno (3.2)
$$
The energy levels of this model are characterized by a
sequence of $N-1$ Bloch-orbital
occupation numbers $\{n_1,n_2,\ldots,n_{N-1}\}$, taking
values 0 or 1, so
$$
E = \sum_{m=1}^{N-1} \epsilon_m(n_m-\half ) ;
\quad e^{iK} = \prod_{m=1}^{N-1}
e^{ik_m n_m}  \eqno (3.3)
$$
where
$$
\epsilon_m = \left ( {2 \pi v_s  \over N^2}\right ) m(m-N)
;\quad k_m = 2\pi m/N   . \eqno (3.4)
$$
All eigenstates have a two-fold degeneracy, corresponding to the
two possible occupation states of the translationally-invariant
Bloch orbital,
which has zero energy.   This is the supersymmetry: all states
are $sl_{1|1}$ doublets, and there is no non-trivial Yangian
extension in this case.   The ground state of $H$ is the state
with maximum occupancy, $\{n_m\}$ = $\{111 \ldots 11\}$, and
the ground state of $-H$ has minimum occupancy $\{000 \ldots 00\}$;
The replacement $1 \leftrightarrow 0 $ in any configuration
maps $E$ to $-E$ and $ K$ to $ \pi(N-1) - K $.

It is a remarkable fact that this set of energy-momentum levels
is the {\it complete} set of levels contained in the general
$SU(m|n)$  exchange model, but in the general case
they have different and {\it much larger} Yangian
multiplicities.   Specializing to the $SU(n|0)$ and $SU(0|n)$ cases,
some of these multiplicities are zero, and the energy level
is absent.  The selection rule for allowed multiplets in the case
$SU(n|0)$ is that
occupation patterns containing a sequence of $n$ or more consecutive
1's are forbidden; in the  $SU(0|n)$ case,
$n$ or more consecutive 0's are forbidden.
Each allowed sequence corresponds
to an irreducible Yangian multiplet.
In general, the spectrum of $H$ with $SU(m|n)$ symmetry is the same
as that of $-H$ with $SU(n|m)$ symmetry.

The above rule was found empirically by examination of
numerical diagonalization results \cite{haldane92}, but I
now specialize to the simplest case, $SU(2|0)$ $\equiv$ $SU(2)$,
where the rule is that occupation
patterns with consecutive 1's are forbidden, and make
contact with the $Y(sl_2)$ representation theory \cite{chari}.

Recall that
each YHWS labeled by $\nu$
has an associated Drinfeld polynomial, $P_{\nu}^+(u)$,
and that $\Delta (u) {\bf T}(u)$ is a polynomial,
with quantum determinant $\Delta (u-h) \Delta (u+h) $.
Thus there is a polynomial $f_{\nu}(u)$ where
$\phi_1^+(u)$, (the eigenvalue of $T^{11}(u)$),
has the form  $f_{\nu}(u)P^+_{\nu}(u+h)$,
$\phi_2^+(u)$ = $f_{\nu}(u)P^+_{\nu}(u)$,
and
$$
f_{\nu}(u) f_{\nu}(u-h) P_{\nu}^+(u+h)P_{\nu}^+(u-h) =
\Delta (u+h) \Delta (u-h) .
\eqno (3.5)
$$
Since the roots of $\Delta (u)$ are an $N$-string, it is easily seen
that $P_{\nu}^+(u)$ must be a factor of $\Delta (u)$, so
$\Delta (u)$ = $P^+_{\nu}(u)g_{\nu}(u)$, where $g_{\nu}(u)$ is
a polynomial with roots at the roots of $\Delta (u)$
{\it not } contained in $P_{\nu}^+(u)$, and
$$
f_{\nu}(u)f_{\nu}(u+h) = g_{\nu}(u-h)g_{\nu}(u+h).
\eqno (3.6a)
$$
This is a polynomial equation with the elementary solution
$$
g(u) = (u-a)(u-a+h) , \quad f(u) = (u-a-h)(u-a+h) . \eqno (3.6b)
$$
Any product
of such solutions is a solution.  This shows that $g_{\nu}(u)$ must
be a product of 2-strings.

We now have the recipe \cite{bernard93} for
constructing the possible Drinfeld polynomials
(in fact, there is one YHWS in the spectrum of the
trigonometric model corresponding to each allowed Drinfeld
polynomial \cite{bernard93}).
First  partition the
$N$-string of  roots of $\Delta(u)$ into 1-strings and 2-strings.
Between each consecutive pair of roots place a 1 if they belong
to the same 2-string, and 0 otherwise.
This gives a binary sequence of length $N-1$, {\it with the constraint
that there are no consecutive $1$'s},  as found empirically.
The 1-strings are the roots
of the Drinfeld polynomial; if an extra 0 is added at each end of
the binary sequence, the 1-strings are located between each consecutive
pair of 0's.  From the Yangian representation theory, we conclude
(in agreement with the  empirical
findings \cite{haldane92}) that a sequence of $M+1$ consecutive
0's represents an independent $S=M/2$ degree of freedom.

There is a simple physical interpretation: each root of the Drinfeld
polynomial represents the presence of a $S=1/2$ ``spinon'' excitation.
A $m$-string of such roots represents $m$ spinons ``in the same orbital''
with a rule that their spin state must be totally symmetric.  If the
total number of spinons present is $N_{sp}$, there are $M$ = $(N-N_{sp})/2$
1's in the occupation pattern, which serve to separate
the $N_{\rm orb}$ =  $M+1$ ``orbitals''
into which the spinons are distributed.   The change in the number
of available ``orbitals'' as the spinon number
is changed obeys
$\Delta N_{\rm orb} / \Delta N_{sp}$ =$-1/2$, which allows
spinons to be identified as excitations
with ``semionic'' fractional statistics, in between
Bose and Fermi statistics \cite{haldane91b}.

If $\{I_i\}$ are the positions of the $M$ non-zero entries in the
length-($N-1$) binary
sequence $\{n_j\}$,  ``spinon orbital occupations''
$\{n_{i\sigma}\}$, $1 \le i \le M+1$, $\sigma $ = $\pm 1/2$,
are given by
$I_{i+1}- I_i$ = 2 +  $n_{i\uparrow} + n_{i\downarrow} $,
with $I_0 \equiv -1 $,
and $I_{M+1} \equiv N+1 $.
We can then
treat the $\{n_{i\sigma}\}$ as independent variables subject
only to the
constraint
$$
N_{sp} \equiv  \sum_{i\sigma} n_{i\sigma}  = N-2M .
\eqno (3.7)
$$
Each distinct configuration $\{n_{i\sigma}\}$ satisfying this constraint
corresponds to an eigenstate of the system.  The energy and momentum
can now  be expressed in terms of the ``spinon orbital occupations'':
it is convenient to relabel $n_{i\sigma}$,  with $ 1 \le i \le M+1 $,
as $n_{k\sigma}$, where $k$ is a crystal
momentum in the range $-k_0 \le k \le k_0$,
$ k_0 $ = $\pi M / N $, then
if
$$ \epsilon (k) = {v_s \over \pi } (k_0^2 - k^2 ) ;
\quad V(k) = v_s (k_0 - |k| ) ; \eqno (3.8)
$$
the spectrum is given by
$$ E = E_{MN}
+ \sum_{k\sigma} \epsilon (k) n_{k\sigma}
+ {1 \over 2 N} \sum_{k\sigma,k'\sigma'}
V(k-k') n_{k\sigma}n_{k'\sigma'} , \eqno (3.9a)
$$
$$
e^{iK} = (-1)^M\prod_{k\sigma} e^{ikn_{k\sigma}} ; \quad J^z_0
= \sum_{k\sigma} \sigma n_{k\sigma}  .
\eqno (3.9b)
$$
Here
$E_{MN}$ =
$\pi v_s (2N(N^2 -1 ) + 4M(M^2+2) - 3MN^2)/6N^2 $.  From this spectrum, it
is straightforward to derive the thermodynamic potentials
 \cite{haldane91a} in the limit $N \rightarrow \infty $, including
a Zeeman coupling $-hJ^z_0$.

The spinon orbitals can
be parameterized by a {\it rapidity} $x$ in the range $ -1 < x < 1$,
(where the spinons in that orbital have velocity $v_sx$),
and have mean occupation numbers
$$
 \bar{n}_{\sigma}(x)
 = \exp -\beta [\varepsilon (x) -\sigma h(1+ \mu (x) ) ] ,
\eqno (3.10)
$$
where $ \varepsilon (x) $ = $(\pi v_s /4)(1 - x^2)$ and
$$
{ \sinh (\beta h \mu (x)  /2 ) \over
 \sinh (\beta h /2 ) }
 = \exp (-\beta  \varepsilon (x) ) .
\eqno (3.11)
$$
This gives an easily-solved
quadratic equation for $\exp( \beta h \mu (x) /2)$.
The free energy per site is given by
$$
-\beta f (\beta , h) =  { 1 \over 2 } \int_{-1}^{1} dx \, \ln
\left ({ \sinh ( \beta h (1 + \mu (x) )/2 )
\over
\sinh (\beta h /2 ) }  \right ).
\eqno (3.12)
$$
In the absence of a magnetic field, the entropy per site
is
$$
s(\beta , h=0) = k_B \int_{-1}^1 dx \left (
\ln [2\cosh \beta \varepsilon (x)] -
\beta \varepsilon (x) \tanh \beta \varepsilon (x) \right ) ,
\eqno (3.13)
$$
which  unexpectedly is {\it even} in $\beta$, so is the same for
the ferromagnetic and antiferromagnetic models.

The identification of spinons as {\it semions}
is supported when the {\it wavefunctions}
of the trigonometric $SU(2)$ model are examined.  A class of polynomial
wavefunctions of the  Calogero-Sutherland model
of a non-relativistic gas with $1/\sin^2$ interactions was
discovered \cite{haldane88,shastry88}
to also give a class of
``fully-spin-polarized spinon gas'' \cite{haldane91a}
wavefunctions of the trigonometric model,
and these are now identified with the YHWS states.
If $Z_i$ (with $(Z_i)^N $ = 1)
are the complex coordinates of the $M$ sites with $\sigma_i$ = $-1/2$,
the wavefunctions
have the form
$$
\Psi_{\nu} (\{Z_i\}) = \Phi_{\nu} (\{Z_i\}) \Psi_0 (\{Z_j\}) ,
\eqno (3.14)
$$
where
$\Phi_{\nu} (\{Z_i\})$ is a symmetric polynomial with
degree $N_{sp}$ in each $Z_j$, and $\Psi_0 (\{Z_j\}) $
is
$$
\Psi_0(\{Z_i\}) = \prod_{j<k} (Z_j-Z_k)^2 \prod_j Z_j  .
\eqno (3.15)
$$
The is essentially the
same as the $m$=2  Laughlin fractional quantum Hall effect state for
bosons \cite{laughlinkalmeyer}, and
for $N_{sp} = 0 $,
this is the unique Yangian singlet state.
The symmetric polynomials $\Phi_{\nu}$ are solutions
of the eigenvalue equation
$$
\left ( \sum_j x_j^2 {\partial^2 \over \partial x_j^2 }
+ \lambda
\sum_{j \ne k} \left ( { x_j^2 \over x_j - x_k } \right ) {\partial \over
\partial x_j } \right ) \Phi^{(\lambda )} (\{x_i\}) =
\mu \Phi^{(\lambda )} (\{x_i\} ) ,
\eqno (3.16)
$$
with $\lambda = 2$.
(The solutions are known in the Mathematical Literature
as the Jack Polynomials \cite{stanley}).
If $\{I_i\}$ are the positions of the non-zero entries in the binary
``occupation number'' sequence, the Taylor series expansion of the
corresponding YHWS wavefunction has the form
$$
\Psi  = \sum_{\{m_i\}} C(\{m_i\} ) \sum_{P}
\left ( \prod _i (Z_{P(i)})^{m_i} \right ) ,
\quad \{m_i\} \le
\{I_i\}
\eqno (3.17)
$$
where $P(i)$ is a permutation, and
where $m_i \le m_{i+1} $, and $\{m_i\} < \{m'_i\}$ means that
$\{m_i\}$ can be reached from $\{m'_i\}$ through a sequence of
pairwise ``squeezing'' operations $m_i$ $ \rightarrow$ $ m_i + 1$,
$m_j$ $\rightarrow$ $ m_j - 1$, with $m_i < m_j-1$, and
$m_k$ unchanged for $k \ne i,j$.

The $N_{sp}$ = 0 Yangian singlet state may also be written in terms of
the azimuthal spin variables $\sigma_i$:
it occurs only for even $N$, and
is the $n$ = 2 case of the $SU(n)$ singlet wavefunction
where the $\sigma_i$ can take one of $n$ ordered values $\{\alpha\}$:
$$
\Psi_0^{(n)} (\{z_i,\sigma_i\}) =
\prod_{i<j} (z_i-z_j)^{\delta (\sigma_i,\sigma_j)}
(i)^{{\rm sgn}(\sigma_i,\sigma_j)}
\prod_{\alpha} \delta (N,nN(\alpha )),
\eqno (3.18)
$$
where
$N(\alpha )$ $\equiv$ $ \sum_i \delta (\alpha ,\sigma_i)$.
Here the $\{z_i\}$ are a complete set of the $N$'th roots of unity.
Note that this state is essentially the same as the wavefunction
for a filled Landau level of $SU(n)$ fermions, and
is hence explicitly $SU(n)$-singlet for {\it arbitrary} $\{z_i\}$.

The $N_{sp}$ = 1 states are also particularly simple: they occur
only for odd $N$ = $2M+1$, and are generated by
$$
\Psi (z;\{Z_j\}) = \prod_i (z- Z_i)
\prod_{i<j} (Z_i - Z_j)^2 \prod_i Z_i .
\eqno (3.19)
$$
Expanding this in powers of $z$ gives a band of Bloch states
with crystal momentum $-\pi/2$ $<$ $K + M\pi$ $ < $ $\pi /2$, so
the spinon band covers {\it half} a Brillouin zone.
The localized spinon wavefunction $\Psi (z;\{Z_i\}) $ is essentially
the quasihole excitation of the $m=2$ bosonic Laughlin state, and
from this perspective, clearly describes a {\it semionic} excitation.
If the parameter $z$ is chosen to be a lattice coordinate  $z_i$,
the localized spinon
wavefunction with $\sigma_i$ = $\sigma$
can be rewritten in terms of spin variables as
$$
 \delta (\sigma,\sigma_i)\Psi_0^{(2)} (\{z_j,\sigma_j; j \ne i \}) .
\eqno (3.20)
$$
This shows that the spinon is {\it completely} localized
on the lattice site, and induces no spin polarization of
its local environment.  While there are $N$ such
fully-localized
spinon states, they are fundamentally non-orthogonal, since
the expansion in orthogonal Bloch states shows there are only
$(N+1)/2$ independent $N_{sp}$=1 states.

To end this section, I discuss the extension of the state-counting
from $SU(2)$ to the general $SU(n|m)$ case.  The Yangian counting rules
for the $SU(2)$ model are very simple, and allowed the thermodynamic
functions to be explicitly obtained \cite{haldane91a} in closed-form in the
thermodynamic limit,
but become
much more complicated in the general $SU(m|n)$ case.
Recently Sutherland and Shastry \cite{sutherland93} showed
how to recover the $SU(2)$  results for
the thermodynamics, and  generalize them to  $SU(m|n)$,
from a strong-coupling
limit of the exchange-generalization of the Calogero-Sutherland
model \cite{polychronakos,cherednik,hahaldane,wadati}.

The spin chain
degrees of freedom are essentially those of the spin-1/2 Fermi gas
{\it with the charge degrees of freedom removed}.  From this viewpoint,
the spinon is  the spin-1/2 fermion with the charge degrees of
freedom factored out, leaving ``half a fermion'', and hence a semion.
I will interpret Sutherland and Shastry's result \cite{sutherland93} as showing
how to ``put back'' charge degrees of freedom into the
spin chain to recover a spectrum with the familiar degeneracies
of the ideal gas with internal degrees of freedom.  This facilitates
the computation \cite{sutherland93} of the thermodynamics in the
general $SU(m|n)$ case.

Let $b^{\dagger}_k$, $k = 1,\ldots N-1$, be a set of harmonic
oscillator creation operators, and add these degrees of freedom
to the spin chain as follows:
$$
H' = \left ( {v \pi \over N^2 } \right )
\left (  \sum_{i<j} { 1-P_{ij} \over \sin^2(\pi(i-j)/N) } +
\sum_{k=1}^{N-1} 2k(N-k) b^{\dagger}_kb_k \right ) ,
\eqno (3.21a)
$$
$$
e^{iK'} = e^{ iK_s } e^{i K_c} ;
 \quad K_c = {2 \pi  \over N}
\left (\sum_{k=1}^{N-1} k b^{\dagger}_k b_k \right ),
\eqno (3.21b)
$$
where $\exp iK_s$ is the spin-chain translation operator, and $P_{ij}$
is the $SU(m|n)$ exchange operator.

Now let $n_{k\alpha}$ = $c^{\dagger}_{k\alpha}c_{k\alpha}$
be  occupation numbers of a set of orbitals for particles
of a bosonic   or fermionic   species $\alpha$, with
$k = 0 , \pm 1, \ldots \pm \infty $,
subject to the constraint that
$$
N = \sum_{k\alpha}n_{k\alpha}
\eqno (3.22)
$$
is fixed.  From the result of Sutherland
and Shastry, the spectrum of $H'$
(with the constraint (3.22)) is identical to that of
$$
H'' = \left ( {v \pi \over N^2 } \right )
\sum_{kk'} \sum_{\alpha\beta} |k-k'| n_{k\alpha}n_{k'\beta} ,
\eqno (3.23a)
$$
$$
K'' =  {2 \pi \over N}\sum_{k\alpha} k n_{k\alpha} ; \quad
J^{\alpha\beta}_0 = \sum_{k}
( c^{\dagger}_{k\alpha}c_{k\beta} - (m+n)^{-1}\delta^{\alpha\beta}) .
\eqno (3.23b)
$$
The spectrum of $H''$ is periodic in
$K''$ $\rightarrow$ $K'' +  2\pi$,
corresponding to a shift $n_{k\alpha} \rightarrow n_{k+1\, \alpha}$
of the occupation number pattern.
The precise statement is that the spectrum
of $H''$ in one period of $K''$ coincides with that of $H'$.
In the thermodynamic limit, the free energy of $H'$
can thus be easily
calculated, as can the  free energy contribution from the extra
oscillator modes; the {\it difference} is the spin-chain
free energy \cite{sutherland93}.  However, this does {\it not}
give a simple
method for identifying the Yangian degeneracies of the {\it discrete}
(finite $N$)
spectrum
of $H$ {\it without} the oscillator modes.\
The explicit expressions for the free energy
are in fact only obtained in the limit $N \rightarrow \infty $, when the
Yangian algebra has degenerated
to its ``classical'' $\widehat{sl_{m|n}}_+$
limit.

\vspace{10mm}
\noindent
{\bf 4.  Dynamical Correlation functions}
\vspace{5mm}

\noindent
Let us consider the state  $S^+_i |0 \rangle $, where
$|0\rangle $ is the Yangian singlet ground state of the trigonometric
chain.   It is a spin-1 state, and is easily seen to be a linear
combination of Yangian highest weight states with $N_{sp}$ = 2.
The action of $S^+_i$ is to remove a down-spin coordinate at site $i$.
Thus $M = N/2 - 1$ and
$$
\Psi (z_i;\{Z_j\}) = \prod_{j=1}^M (z_i -Z_j)^2 \Psi_0(\{Z_j\}) .
\eqno (4.1)
$$
The polynomial prefactor has degree 2, confirming that
this state is composed purely of two-spinon eigenstates.
Thus if we wish to compute the dynamical correlation function
$$
C(i-j,t-t') = \langle 0 | S^-_i(t)S^+_j(t') | 0 \rangle ,
\eqno (4.2)
$$
the only
intermediate states that contribute to its spectral function
are the $N_{sp}$=2 YHWS.

Recently, it has become possible to compute certain
dynamical correlation functions
of the Calogero-Sutherland model \cite{calogero,sutherland}
$$
H = \sum_{i=1}^N { p_i^2 \over 2m}
+  {\hbar ^2 \over m }
\sum_{i < j} { \lambda(\lambda -1) \over d(x_i - x_j)^2 } ,
\eqno (4.3)
$$
with $d(x)$ = $(L/\pi )\sin ( \pi x /L)$, in the thermodynamic limit
at fixed density $\rho$ = $N/L$.
The eigenfunctions of the CSM have the form
$$
\Phi^{(\lambda)} (\{Z_i\}) \Psi_J(\{Z_i\}); \quad \Psi_J =
\prod_{i<j} (Z_i-Z_j)^{\lambda}\prod_i (Z_i)^J \quad (\lambda \ge 0 )
\eqno (4.4)
$$
where $\{\Psi_J\}$ is the family of states that includes
the ground state and Galilean boosts of it,
$Z_i$ = $\exp 2\pi i x_i/L $, and
$\Phi^{(\lambda)} (\{Z_i\} )$ is a symmetric polynomial
solution of (3.16).
While the apparent statistics can be modified with a singular
gauge transformation,
the ``natural'' statistics of this model are, from (4.4),
evidently {\it fractional},
with statistical parameter $\theta$ = $ \pi \lambda $
({\it i.e.}, particles carry charge 1 and flux $\pi \lambda $).
The elementary excitations are particles with velocities greater
than the speed of sound $v_s$ = $\pi \hbar \lambda \rho /m $,
and {\it holes} with velocities less than $v_s$.   The holes
carry charge $-1/\lambda $ and flux $-\pi $, and have statistical
parameter $\theta_h$ = $\pi / \lambda $.

The calculations can be carried out at one of three
special couplings (only two of which are non-trivial), and  involve a
mapping to the $N \rightarrow \infty $ limit of
a Gaussian dynamical $N \times N$
matrix-model, with either orthogonal ($\lambda$ = 1/2),
unitary ($\lambda$ = 1), or symplectic ($\lambda$ = 2) symmetry.  This reduces
the problem
to a complicated, but tractable Gaussian problem \cite{simonsetal}.

At integer couplings $\lambda$ = $q$, the natural particle statistics
are Bose (even $q$) or Fermi (odd $q$).  The ground state is
$\Psi_{J}$ with $J$ = $-q(N-1)/2$.  The wavefunction of
the state $\Psi (x) | J \rangle $ is
$$
\prod_i (z-Z_i)^q \Psi_J(\{Z_i\}),\quad z = \exp(2\pi i x /L ),
\eqno (4.5)
$$
which is composed only of eigenstates with just $q$ hole
excitations, so the
spectral function of the retarded single-particle Greens function
will only contains contributions from intermediate states
of that type.
This correlation function will thus have the
form
$$
\langle 0 | \Psi^{\dagger}(x,t) \Psi (0,0) | 0 \rangle
=
\rho \left ( \prod_{i=1}^{q} \int_{-v_s}^{v_s} \! \! \! dv_i  \right )
|f_q(\{v_i\})|^2 e^{i(Px-Et)} ,
\eqno (4.6a)
$$
$$
P =  \sum_i m_hv_i  , \quad  E =
\sum_i \half m_h v_i^2  ,\quad m_h = -m/q ,
\eqno (4.6b)
$$
where $f_q(\{ v_i\}) $ is a {\it form factor} that must be calculated.

Using the mapping to the symplectic matrix
model, it has been possible  \cite{haldane93} to calculate this  form factor
for $\lambda $ = $q$ = 2.   The result is remarkably simple:
$$
|f_2(v_1,v_2)|^2 =  {1  \over 8v_s}  \left (
{ (v_1 - v_2)^2 \over (v^2_s - v^2_1) (v^2_s - v_2^2) }
\right ) ^{\half} .
\eqno (4.7)
$$
The significance of this in the context of the $SU(2)$ trigonometric
Haldane-Shastry chain is that the YHWS wavefunctions
with $Z_i$ = $ \exp  ( 2\pi i x_i /L )$ are also
eigenstates of the $\lambda = 2$ CSM, and  the
matrix elements of $S^+_i(t)$ between two YHWS are equivalent
to those of  $\Psi (x_i,t)$ between the corresponding CSM states.
{\it The CSM result can immediately be translated to the
results for the spin chain in the {\it rational} limit, where
the Yangian symmetry algebra becomes classical.}

While the calculation \cite{haldane93}
is complicated and indirect, the simplicity
of the result suggests there should be a simple derivation, not
based on the ``accident'' of
a mapping to a Gaussian matrix model.  It is very tempting
to {\it conjecture} the extension of the result to the $SU(n)$ chain,
for which no such mapping is known.  Examination of the action
of the operator $c^{\dagger}_{i\alpha}c_{i\beta}$ on the singlet
ground state  of the $SU(n)$ chain
leads to the conclusion that it produces a two-parameter
family of YHWS states with one spinon (of ``color'' $\bar{\beta}$)
moving with rapidity  $x_1$, and
a complex of $n-1$ spinons (with net ``color'' $\alpha$) moving
together (but not as a bound state) with rapidity
$x_2$.
In the singlet ground state of the rational
$SU(n|0)$ chain
state, the correlation function
$\langle 0 | X^{\alpha\beta}_j(t)X^{\gamma\delta}_0(0) | 0 \rangle$,
where $X^{\alpha\beta}_i$ $\equiv$ $c^{\dagger}_{i\alpha}c_{i\beta}$,
will thus  have the form
$$
{ (-1)^j \over n} \delta^{\alpha\delta}\delta^{\gamma\beta} C(j,t)
+
{(-1)^j \over n^2 }
\delta^{\alpha\beta}\delta^{\gamma\delta} (1 - C(j,t)) ,
\eqno (4.8)
$$
where
$$
C(m,t) = \frac {1}{4}  \int_{-1}^{1} dx_1 \int_{-1}^{1} dx_2 \,
|F_n( x_1,x_2)|^2
\left ( e^{i(q(x_1)m-\epsilon(x_1)t)} \right )
\left ( e^{i(q(x_2)m-\epsilon(x_2)t)} \right )^{n-1},
\eqno (4.9a)
$$
$$
q(x) =
{\pi x \over n } ;\quad \epsilon (x) = {\pi v_s \over 2n }(1 - x^2 ) .
\eqno (4.9b)
$$
Here $F_n(x_1,x_2)$ is the form factor that must be found.

The asymptotic form of the correlations are straightforward to compute
from bosonization (they are free-fermion correlations with
the factor coming from charge degrees of freedom divided out),
or  from conformal field theory \cite{kawakami},
and can be fit to a simple {\it Ansatz} based on (4.7).
I therefore present the {\it conjecture} for the form factor of the
rational $SU(n|0)$ chain
(which is a  rigorous result for $n$=2) :
$$
|F_n(x_1,x_2)|^2 = A_n \left ( { 4 (x_1-x_2)^2 \over (1-x_1^2)
 (1-x_2^2) } \right )^{1/n}
, \quad A_n = {
\left ( {n-1 \over n} \right ) \Gamma \left ( {n+1 \over n} \right )
\over
\Gamma \left ( {n-1 \over n} \right ) \Gamma \left ( {n+2 \over n} \right )}.
\eqno (4.10)
$$
It has been verified \cite{unpub}
that this form fits the numerically computed
static structure factor of
the trigonometric $SU(3)$ chain with $N \le 18 $ extremely well, with
very small finite-size corrections, leaving no doubt that this
conjecture for $n > 2$ is indeed correct.

As another test of the reliability of generalizing
(4.7) ``by conjecture'', based on its remarkable
simplicity, the same type of arguments
can be used to obtain the form factor for the retarded Greens function
(4.6) of the CSM at a general integer coupling $\lambda = q$.  In this case,
the states contributing to the spectral function have
$q$ hole excitations with independent velocities.  The resulting
conjecture  is
$$
|f_q(\{v_i\})|^2 ={1 \over 2 v_s } B_q
\left ( \prod_{i=1}^q (v_s^2 - v_i^2) \right )^{-1 + 1/q}
\left ( \prod_{i<j} (v_i - v_j)^2 \right )^{1/q} ,
\eqno (4.11a)
$$
where the multi-dimensional integral fixing the normalization
$B_q$ is hard to do.
I have just learned of recent work by Forrester \cite{forrester93}, who
calculates the equal-time limit of this correlation function
using quite different
methods based on the celebrated {\it Selberg trace formula}.
His expression {\it is precisely the equal-time limit of the formula
conjectured here!}  Moreover, Forrester's
result provides the normalization as
$$
B_q =  \prod_{j=1}^q \left ( {\Gamma \left ( {1+ q \over q} \right )
\over
\left ( \Gamma \left ( { j \over q } \right )\right ) ^2  } \right ) .
\eqno (4.11b)
$$
It is striking that the form factors are essentially
Laughlin-type wavefunctions for
{\it anyons} with the same statistics as the holes,
now  with the {\it rapidities} as coordinates.

The remarkable property
of the rational and trigonometric models is that the {\it local}
operators such as the spin on a given site act on the ground
state to produce only a very restricted class of
excitations.  There is a general selection rule,
verified empirically \cite{talstra},
that the local spin operator $\vec{S}_i$ acting
on any eigenstate cannot change the spinon number
$N_{sp}$ by more than $\pm 2 $.
Such properties are very reminiscent of an ideal gas, and
the most natural interpretation of the trigonometric and rational
models is as generalizations of the ideal gas Fock-space
structure to non-trivial
statistics.

\vspace{10mm}
\noindent
{\bf 5.  Conclusion }

\vspace{5mm}
\noindent
I have attempted to present, from my perspective, the main results
associated with the still-unfolding properties of the Haldane-Shastry
spin chain model, stressing the simplest $SU(2)$ or $S=1/2$ model.
I have clearly omitted
many aspects, and
there is clearly much more yet to emerge.

A direct algebraic treatment in the rational limit, when the
``quantum group'' becomes classical, would
be particularly desirable.  It can be no accident that this
is the limit in which all the explicit results are obtained, but
(disappointingly)
so far a direct use of its algebraic properties
such as the infinite $\widehat{sl_2}_+$ symmetry
not yet been made.  An algebraic construction
of the highest weight eigenstates (fully spin-polarized spinon gas states)
in terms of particle creation operators (``vertex operators'')
acting on the vacuum is also needed; this would closely parallel
a similar treatment needed for the Calogero-Sutherland model.

Other possible lines of investigation are whether a ``quantum
deformation'' $\widehat{sl_2}_+$ $\rightarrow$ $Y(sl_2)$ can be used to
calculate the hyperbolic model form factors, and what is the origin of
integrability of the elliptic models.
There is clearly much more work to be done!

I thank B.\ S.\ Shastry for bringing Ref.\cite{forrester93}
to my attention, and wish to express my
appreciation to the Taniguchi Foundation for
its vision in making  Symposia such as this one
possible.

\vspace{10mm}


\end{document}